\documentclass[%
reprint,
 amsmath,amssymb,
]{revtex4-1}

\usepackage{graphicx}
\usepackage{dcolumn}
\usepackage{bm}

\begin{document}

\preprint{APS/123-QED}

\title{Interband interaction between Ge states and surface resonance band of Pb on Ge(001)}

\author{Tomohiro Sakata}
 \email{s-tomohiro@ms.naist.jp}
\author{Sakura N. Takeda}
\author{Kousuke Kitagawa}
\author{Haruka Kumeda}
\author{Kazuki Kokui}
\author{Hiroshi Daimon}

\affiliation{
 Graduate School of Materials Science, Nara Institute of Science and Technology, Nara 630-0192, Japan
}

\begin{abstract}
We investigate the valence band structure of Pb on Ge(001) by Angle-Resolved Photoelectron Spectroscopy. Three Ge bands, G1, G2, and G3, were observed on Ge(001) 2$\times$1 clean surface. In addition to these three bands, a forth band (R band) is found in the 2 ML of Pb coverage. The R band continues to appear even when the surface superstructure changed. The position of the R band does not depend on Pb coverage. These results indicate that the R band derives from Ge subsurface states known as surface resonance states. Furthermore, the effective mass of G3 is significantly reduced when this forth band exists. We found that this reduction of the G3 effective mass was explained by the interaction of the G3 and the surface resonance band. Consequently, the surface resonance band penetrates the Ge subsurface region affecting the Ge bulk states. We observed the hybridization between Ge states and the surface resonance states induced by Pb adsorption.  
\end{abstract}

\pacs{Valid PACS appear here}
\maketitle
\pagestyle{empty}

\section{\label{sec:level1}Introduction}
The subsurface electronic structure of semiconductors contributes to performance of MOSFETs. For higher performance devices, it is necessary for the carrier to have a lighter effective mass  within the subsurface region of the  semiconductor.  Semiconductor subsurface states are not only  different from  bulk states, but are also dramatically changed by a small amount of adsorption. In the case of silicon, it has been observed that a hole subband structure  is formed by confinement within the p-type inversion layer \cite{PhysRevLett.82.035318}. That means that the Si band structure during operation of a MOSFET is more complicated than the Si bulk band structure. Furthermore, it was reported that the Si parabolic band is changed into liner dispersion due to interaction with surface states caused by Pb deposition\cite{PhysRevLett.104.246803}. This indicates that the electronic structure within the semiconductor subsurface region is tuned by specific surface states. 

Among semiconductor, Germanium (Ge) is a promising material for next generation semiconductor devices because if its higher electron and hole mobility compared with silicon\cite{toriumi2009opportunities}.  Ge  has specific states which are localized within the subsurface region. For example, in the Pb-adsorbed on Ge(111) system, a fourth band was observed within a SOS-induced gap even though the Ge valence band consists of only three bands (HH; Heavy Hole, LH; Light Hole, and SO; Split-Off)\cite{tang2010bilayer}. This fourth band derives from surface resonance (SR) states within the Ge subsurface region. This SR states could  be seen in the different adsorption on Ge(111) system.   This SR states penetrates through the subsurface region mixes with the bulk states. Furthermore, in Bi, Br, and Tl-adsorbed Ge(111) systems, a Rashba-type spin-splitting band which localized in the subsurface region was obtained\cite{PhysRevB.82.201307}. These results suggest that Ge electronic states within the subsurface region can be tuned by adsorption onto Germanium. Therefore, understanding the electronic structure of metal/Germanium within the subsurface region is important information for tuning Ge electronic structures within the subsurface region. Thus, we investigated Pb adsorption effect on the Ge(001) band structure by Angle-Resolved Photoelectron Spectroscopy. 
In this paper, we suggest that Ge band dispersion is tuned by hybridization with surface resonance states formed by Pb adsorption onto Ge(001). As a result of this hybridization, the effective mass of the Ge band is lighter than before hybridization. Our results suggest Ge band engineering technology within the subsurface region by means of the hybridization with surface resonance states.

In Pb-adsorbed Ge(001), a Pb thin film can be formed onto Ge(001) with layer by layer at low temperature\cite{PhysRevLett.79.1527}. Pb quantum well states (QWS) and the coupling between the Pb film and Ge(001) substrate are reported by Chen $\it{et. al.}$\cite{PhysRevB.84.205401}. 
At the preparation temperature below 300 degree, 2x2, c(8$\times$4), $\begin{pmatrix} 2 & 1 \\ 0 & 3 \end{pmatrix}$, and c(8$\times$4)i (incommensurate phase) superstructure were obtained \cite{falkenberg1997lead}. In these low coverage phases (below 1.5 ML) like 2x2 and c(8$\times$4), Pb asymmetric dimer covalently bonded to the Ge substrate was formed. The Ge dimer disrupts in these superstructures. In contrast to these low coverage phases, the distorted Pb(111) overlayers is formed on top of the dimerized Ge substrate in high coverage phases(above 1.5 ML) like $\begin{pmatrix} 2 & 1 \\ 0 & 3 \end{pmatrix}$ and c(8$\times$4)i \cite{falkenberg1997lead}\cite{bunk2001ge}.
Even though the atomic structure of Pb/Ge(001) has been investigated with STM\cite{PhysRevB.51.7571}, LEED\cite{zhang1993surface}, and calculation\cite{PhysRevB.64.035304}, the electronic structure of Pb/Ge(001) has yet to be revealed at low Pb coverage.

\section{\label{sec:level1}Experiment}
We used p-type Ge(001) wafers(resistivity; 0.01-0.03 $\Omega$cm). A combination method of wet treatment and annealing in UHV chamber was used, since Ar sputtering for clean surface induces surface roughness\cite{Ar1}. 
First, the sample was dipped into RCA solution to remove contamination, and then into HF solution for hydrogen termination. After that, this sample was immediately loaded into the UHV chamber system(less than 3$\times$10$^{-10}$ Torr)\cite{yamatani2007total}. The sample degassing and flashing annealing were carried out at 400 $^\circ$C and 700 $^\circ$C to obtain the clean  Ge(001) surface 2$\times$1. Pb was deposited onto the clean Ge(001) surface at RT. The clean Ge(001) 2$\times$1 and the superstructure of Pb/Ge(001) were observed by RHEED.
Pb was deposited onto the clean Ge(001) surface at RT.   ARPES measurement was performed using a SCIENTA analyzer (SES2002) with HeI$ \alpha $(21.2 eV) at RT.
\section{\label{sec:level1}Results and Discussion}

\subsection{\label{sec:level2}RHEED Observation}
\begin{figure}[ht]
 \begin{center}
  \includegraphics[width=8cm,clip]{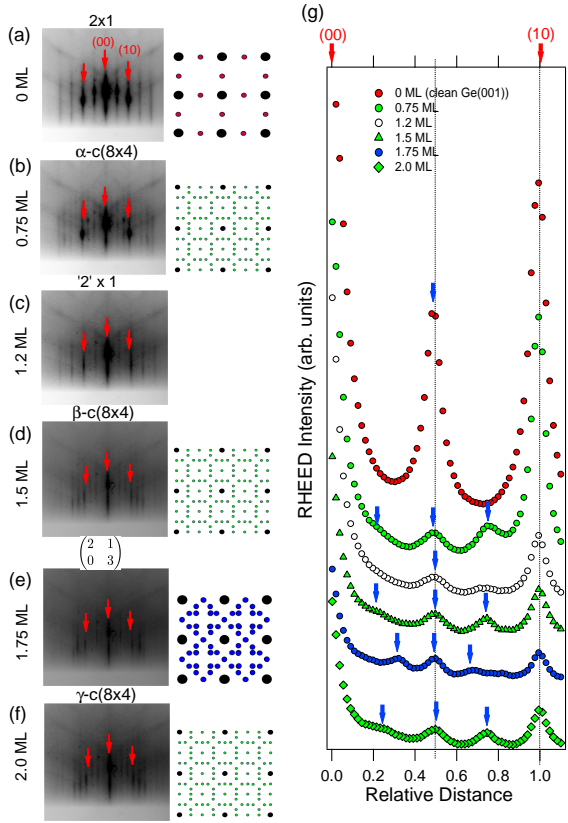}
  \caption{ Observed RHEED pattern of 0th Laue zone(left) and reciprocal spots(right) are shown in (a) at 0 ML, (b) at 0.75 ML, (c) at 1.2 ML, (d) at 1.5 ML, (e) at 1.75 ML, and (f) at 2.0 ML. (g)RHEED spot intensity profile between (00) and (10) fundamental spot. Red and blue arrows are indicated fundamental spots from Ge bulk unit cell and superstructure spots, respectively.}
  \label{RHEED}
 \end{center}
\end{figure}

The surface structure of Pb adsorbed- Ge(001) has been investigated by other group using STM and calculations\cite{falkenberg1997lead}\cite{PhysRevB.51.7571}. According to these previous studies, there are 2$\times$2, c(8$\times$4), $\begin{pmatrix} 2 & 1 \\ 0 & 3 \end{pmatrix}$ and c(8$\times$4)i phases at RT\cite{falkenberg1997lead}. Our RHEED results are shown in Fig.\ref{RHEED} together with the corresponding reciprocal lattice. Fig.\ref{RHEED}(g) also shows the intensity profiles between the (00) spot and (10) spot of RHEED patterns (a)-(f). The 1/2 spots were clearly observed in clean Ge(001)2$\times$1(see Fig.\ref{RHEED}(a) and (g)). At 0.75 ML, four-times periodicity spots derived from c(8$\times$4) appeared at 0th Laue zone in the RHEED pattern(see Fig.\ref{RHEED}(b) and (g)). Therefore this first c(8$\times$4) corresponds to the phase diagram reported by G. Falkenberg $\it{et. al.}$\cite{falkenberg1997lead}. However, we found  that the intensity of these spots relating to c(8$\times$4) decreased above 0.75 ML and completely vanished at 1.2 ML, as shown in Fig.\ref{RHEED}(c) and (g). After that, c(8$\times$4) spots clearly appeared at 1.5 ML again, indicating that c(8$\times$4) obtained at 1.5 ML is different from the first  c(8$\times$4) at 0.75 ML. At 1.75 ML, Three-times periodicity spots  were observed in addition to some spots derived from $\begin{pmatrix} 2 & 1 \\ 0 & 3 \end{pmatrix}$(see Fig.\ref{RHEED}(d) and (f)). Furthermore, Four-times periodicity spots were obtained again at 2.0 ML (see Fig.\ref{RHEED}(e) and (f)). This structure corresponds to the c(8$\times$4)i phase reported by G. Falkenberg $\it{et. al.}$\cite{falkenberg1997lead}. In our experiment,  c(8$\times$4), $\begin{pmatrix} 2 & 1 \\ 0 & 3 \end{pmatrix}$ and c(8$\times$4)i phases were also observed at 0.75 ML, 1.75 ML, and 2.0 ML, respectively. In addition to these reported superstructures, a new c(8$\times$4) was found at 1.5 ML. In this paper we refer to these as $ \alpha $-c(8$\times$4) at 0.75 ML, $ \beta $-c(8$\times$4) at 1.50 ML, and $ \gamma $-c(8$\times$4) at 2.0 ML.

\subsection{\label{sec:level2}ARPES study}
\begin{figure*}[!t]
 \begin{center}
  \includegraphics[width=13cm,clip]{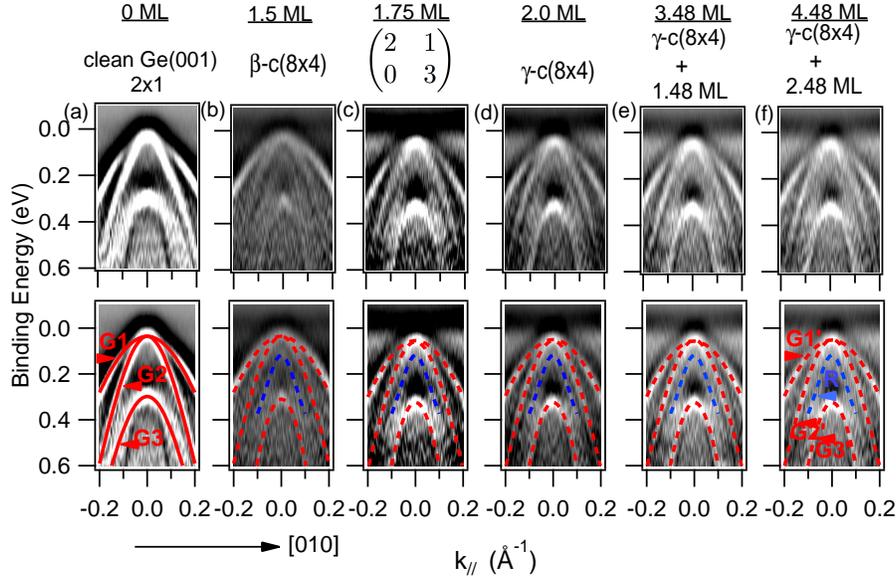}
  \caption{Top panel; Second derivative of ARPES spectrum on (a)the clean Ge(001), (b)1.5 ML, (c)1.75 ML, (d)2.0 ML, (e)3.48 ML, and (f)4.48 ML of Pb on Ge(001). Bottom panel; Band dispersion obtained by fitting line on ARPES data. Fitting lines obtained by Eq.\ref{pade} indicated by the solid line for clean Ge(001) and the dotted line for Pb-adsorbed Ge(001) }
  \label{ARPES1}
 \end{center}
\end{figure*}

ARPES results for the clean Ge(001) surface and Pb-adsorbed Ge(001) are shown in Fig. \ref{ARPES1}. ARPES measurements was performed with [010] direction. In the band dispersion of clean Ge(001) (see Fig. \ref{ARPES1}(a)), three band structures, G1, G2, and G3 are clearly shown. It has been reported experimentally \cite{PhysRevB.72.241308} and theoretically \cite{PhysRevLett.103.189701} that the top of the valence band structure derives from Ge bulk states. Energy separation between G2 and G3 was 0.26 eV at $\Gamma$ point. This value is in roughly good agreement with the spin-orbit gap of Germanium ($\Delta_{SO}\simeq$0.29 eV). H. Seo {\it et al.} suggested that the G1 band derives from the surface resonance states\cite{seo2014critical}.
We also observed three Ge bands G'1, G'2, and G'3 after Pb deposition (see Fig. \ref{ARPES1}(c),(d)). It is striking that a fourth band was observed on the inside of the G2 band after Pb deposition. We refer to this as the R band. In order to compare the band dispersion for the clean Ge(001) surface with the Pb-adsorbed Ge(001) surface and determine the effective mass of each bands, we obtained the fitting line by pad\'e function(shown in Eq.(\ref{pade})) (see Fig. \ref{ARPES1}(b), (d)). 
\begin{eqnarray}
E_{(k_{//})}=E_{(0)}+\frac{\hbar^2 {k_{//}}^2}{2m^*} \frac{1+a{k_{//}}^2}{1+b{k_{//}}^2} 
\label{pade}
\end{eqnarray}

\begin{figure}[ht]
 \begin{center}
  \includegraphics[width=8cm,clip]{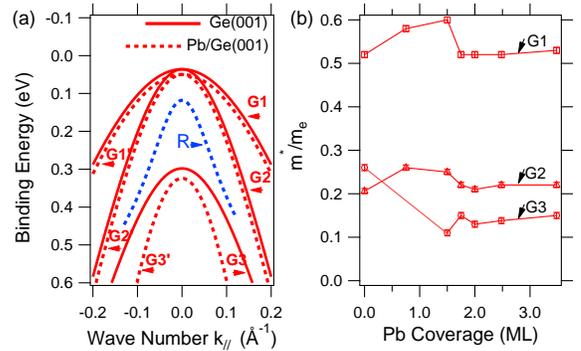}
  \caption{(a)Band dispersion obtained by the fitting lines of Ge(001) and Pb 2 ML adsorbed-Ge(001) (b)Pb coverage dependence of effective mass of G'3 band}
  \label{em}
 \end{center}
\end{figure}

$E_{(0)}$ is the energy level at $\Gamma$ point. Fitting parameters, a, b, and the zone center effective mass ($m^*$) were determined by photoelectron peak positions. Previous studies show this equation is effective for fitting non-parabolic bands\cite{tang2010bilayer}. As shown in Fig. \ref{em} (a), G1 and G2 band dispersions are almost the same with the band dispersion after Pb deposition. On the other hand, the G'3 dispersion becomes shaper than that of clean Ge(001) in Fig.\ref{ARPES1}(e). Furthermore, this band shifted by 0.03 eV towards higher binding energy. This amount of shift is bigger than the other bands because HH and LH bands shifted by 0.01 eV at $\Gamma$ point.  Pb coverage dependence of the effective mass of G'3 band is shown in Fig.\ref{em}(b). In 0.75 ML, $ \alpha $-c(8$\times$4), we could not obtain enough intensity to get a photoelectron peak from the G'3 band. The effective mass of G3 was reduced above 1.5 ML of Pb coverage. In terms of effective mass, it becomes halved by Pb deposition. 
In the next section, we focus on the following two questions: What is the origin of the R band? Why did the G'3 band dispersion change?

\subsection{\label{sec:level2}Origin of R band}
\begin{figure}
 \begin{center}
  \includegraphics[width=6cm,clip]{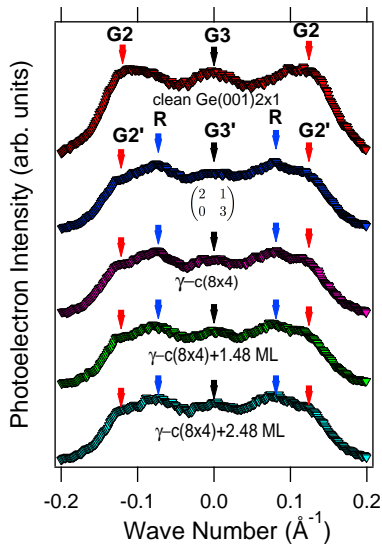}
  \caption{MDC(Momentum distribution curve) at 0.26 eV from VBM(Valence band maximum) at each superstructure. Solid white, blue and red arrow means peak position of G'2, G'3 and R band respectively.}
  \label{mdc1}
 \end{center}
\end{figure}

In order to elucidate the origin of the R band, Pb coverage-dependence of the MDC curve at 0.26 eV from Valence band maximum (VBM) is shown in Fig. \ref{mdc1}. We obtained a different surface superstructure at each Pb coverage (see Fig. \ref{RHEED}) until 2.0 ML. 
Above 2.0 ML, the c(8$\times$4) structure could always be seen. MDC spectra were fitted by Lorenz function to get the peaks and peak positions. These peak positions are indicated by arrows in Fig. \ref{mdc1}. There are three possible states as the conceivable origin of the R band. The first is surface states. It should be localized in the surface. This states should vanish with a change in surface superstructure because of its dependence on surface structure. In these MDC spectra of Pb-adsorbed Ge(001), the R band can be seen not only for $\begin{pmatrix} 2 & 1 \\ 0 & 3 \end{pmatrix}$, but also $ \gamma $-c(8$\times$4). It was found by the result of Fig. \ref{mdc1} that even though the superstructure of Pb on Ge(001) is changed, the peak deriving from the R band still appear. It means that the R band does not depend on superstructure. The second possible origin of the R band is the Pb quantum well states. These states are confined within the Pb thin film. Therefore, this band position should be changed by Pb coverage because confined width becomes wider with increasing Pb coverage. The position of the R band does not change despite an increase in Pb coverage (Fig. \ref{mdc1}). This feature indicates that the origin of the R band is not Pb quantum well states. The third possibility is surface resonance states. These states should be localized within the Ge subsurface region. The existence of surface resonance states does not depend on the change of the superstructure owing to penetration through the Ge subsurface region. The position of the band does not changed with an increase in Pb coverage. We concluded that the origin of the R band observed in the ARPES data above 1.5 ML of Pb derives from the surface resonance states. 
The observed R band dispersion does not changed above $\begin{pmatrix} 2 & 1 \\ 0 & 3 \end{pmatrix}$ superstructure. 
In high coverage phases like $\begin{pmatrix} 2 & 1 \\ 0 & 3 \end{pmatrix}$ and  c(8$\times$4)i, the Ge dimer is completed under the Pb overlayer \cite{falkenberg1997lead}\cite{bunk2001ge}.  
We suggested that this R band induced by Pb adsorption localize from the interface between Pb overlayer and  Ge substrate to the Ge subsurface region. 
Thus, the reason why the R band dispersion does not change is that the Ge dimer structure under the Pb overlayer keeps above $\begin{pmatrix} 2 & 1 \\ 0 & 3 \end{pmatrix}$.

\subsection{\label{sec:level2}Interband interaction between SR states and Ge states}
We found that the G3 band dispersion(effective mass) was dramatically changed after Pb deposition (shown in Fig.\ref{em}). This change in the G3 band dispersion is caused by the 1.5 ML of Pb coverage. The R band also appeared at this Pb coverage.
Therefore, the change in G3 band dispersion correlates with the appearance of the R band. We assumed that the G3 band dispersion was changed by the interband interaction with the R band. When the G3 band is hybridized with the R band, hybridized dispersions E$_{\pm}$ is obtained by Eq.(\ref{interaction1}). 

\begin{eqnarray}
E_{\pm}=\frac{1}{2}(E_{\rm{R}}+E_{\rm{G3}})\pm \frac{1}{2} \sqrt{(E_{\rm{R}}-E_{\rm{G3}})^2 +4V^2} 
\label{interaction1}
\end{eqnarray}

\begin{figure}
 \begin{center}
  \includegraphics[width=9cm,clip]{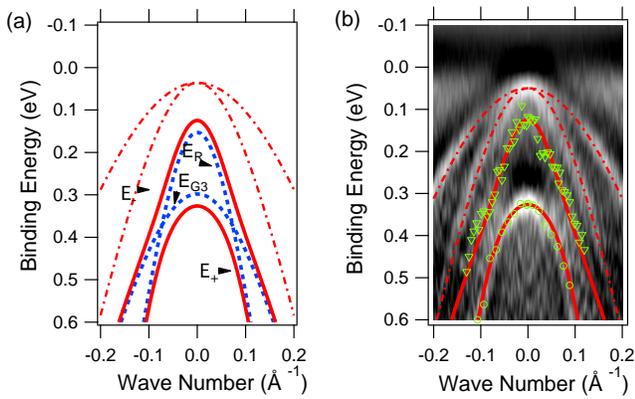}
  \caption{(a)Band structure with considering inter band interaction. Energy dispersions without hybridization, $E_{\rm{G3}}$ and $E_{\rm{R}}$ are shown in dashed blue line. Hybridized energy dispersions, $E_{\rm{+}}$ and $E_{\rm{-}}$ are indicated by solid red line. (b)Comparison of hybridized dispersion with photoelectron peak position of G'3 and R band indicated by the open circle and triangle on intensity map.}
  \label{ARPES2}
 \end{center}
\end{figure}

Here, $E_{\rm{G3}}$ represents the G3 band dispersion for clean Ge(001) (see \ref{ARPES1}(b)). $E_{\rm{R}}$ is obtained by fitting photoelectron peaks derived from the R band within a narrow momentum range($\pm$0.05 $\AA$$^{-1}$). Hybridization energy $V$ (= 0.08 eV) is a fitting parameter. The obtained energy dispersions are shown in \ref{ARPES2}(a). For G3 band dispersion, $E_{\rm{G3}}$ of clean Ge(001) becomes sharper by the hybridization with the R band.  
In addition, hybridized energy dispersions, $E_{\rm{+}}$ and $E_{\rm{-}}$, are in good agreement with the photoelectron peaks derived from the R band and G'3 band as shown in Fig.\ref{ARPES2}(b). Therefore, the change in G3 band dispersion can be explained by the interband interaction with the R band. These results indicate that the G3 band dispersion in Ge subsurface region is hybridized with the R band after Pb deposition. We found that the origin of the R band is the surface resonance states within the Ge subsurface region. Surface resonance states have maximum amplitude of wave function at the surface and penetrate through the bulk by mixing with bulk states. Therefore, we assume that these surface resonance states interact easily with Ge bulk states within the subsurface region because these surface resonance states are located close to the bulk states in terms of real space and $E$-$k$ space.

\section{\label{sec:level1}Conclusion}
In summary, by ARPES we revealed the hybridization effect of the surface resonance band in the Ge bulk states within subsurface region. Three Ge bands, (G1, G2, and G3 band) were observed in clean Ge(001). In contest, a forth band, (R band) was observed above 1.5 ML of Pb coverage.  
Although the surface superstructure changes with the different 
Pb coverage, this R band still appeared. In addition, the position  and shape of the R band  does not depend on Pb coverage. Therefore, we found that the R band derives from surface resonance states localized within  the Ge subsurface region. One of the Ge bands was dramatically changed by Pb deposition at the same time as the appearance of R band. This change in band dispersion can be explained by the interband interaction between the G3 band and R band with the hybridization energy of 0.08 eV. We experimentally observed that surface resonance band induced by Pb deposition affects Ge states. Consequently, the effective mass of the G3 band reduced by half, due to this interaction with the R band. This result indicates that Ge subsurface states can be tuned by monolayer metal deposition. It was recently reported that a high quality interface between Ge and Ge oxide and higher mobility were obtained by a small amount of metals. We assume that this interaction can be a new band engineering technique to make Ge MOS device.

\bibliography{apssamp}

\end{document}